\documentclass[12pt]{article}
\pdfoutput=1

\usepackage{subcaption} % for subfigures

\usepackage[utf8]{inputenc}
\usepackage[tbtags]{amsmath} 				% split tags on lower right
\usepackage{amssymb, mathrsfs}			% for math script fonts
\usepackage{physics}
\usepackage{graphicx}			% for images
\usepackage{tensor}				% for tensors
\usepackage[makeroom]{cancel}	% for cancel
\usepackage{bm}								% for bold Greek
\usepackage[
      colorlinks=true,
      linkcolor=blue,
      urlcolor=blue,
      filecolor=black,
      citecolor=red,
      pdfstartview=FitV,
      pdftitle={},
      pdfauthor={Michael Gutperle, Charlie Hultgreen-Mena},
      bookmarksopen=true
      ]{hyperref}
\usepackage{calc}
\usepackage{sectsty}
\usepackage{dsfont}
\usepackage{cite}

\newlength\myheight
\newlength\mydepth
\settototalheight\myheight{Xygp}
\def\ep{\varepsilon}

\DeclareMathOperator{\diag}{diag}

\DeclareMathOperator{\SO}{SO}

\DeclareMathOperator{\OSp}{OSp}

\newcommand{\AdS}{\text{AdS}}
\newcommand{\cA}{\mathcal A}
\newcommand{\cB}{\mathcal B}

\newcommand{\cK}{\mathcal K}

\newcommand{\cP}{\mathcal P}
\newcommand{\cN}{\mathcal N}
\newcommand{\cM}{\mathcal M}

\newcommand{\cV}{\mathcal V}
\newcommand{\cQ}{\mathcal Q}
\newcommand{\cL}{\mathcal L}

\sectionfont{\large}

\renewcommand{\thefootnote}{\fnsymbol{footnote}}
\renewcommand{\thanks}[1]{\footnote{#1}}
\newcommand{\starttext}{
\setcounter{footnote}{0}
\renewcommand{\thefootnote}{\arabic{footnote}}}
\renewcommand{\epsilon}{\varepsilon}	% use varepsilon
\numberwithin{equation}{section} 		% number equations by section

\numberwithin{equation}{section}

\long\def\symbolfootnote[#1]#2{\begingroup%
\def\thefootnote{\fnsymbol{footnote}}\footnote[#1]{#2}\endgroup}

\begin{document}
\setlength{\baselineskip}{16pt}

\starttext
\setcounter{footnote}{0}

\begin{flushright}
\today
\end{flushright}

\bigskip

\begin{center}

{\Large \bf  Janus and RG interfaces in three-dimensional  gauged supergravity II: General $\alpha$ }

\vskip 0.4in

{\large Michael Gutperle  and  Charlie Hultgreen-Mena }

\vskip 0.2in

{\sl Mani L.~Bhaumik Institute for Theoretical Physics}\\
{\sl Department of Physics and Astronomy }\\
{\sl University of California, Los Angeles, CA 90095, USA}

\end{center}
 
\bigskip
 
\begin{abstract}
\setlength{\baselineskip}{16pt}

In this paper, we  continue the study of Janus and RG-flow interfaces in three dimensional supergravity continuing the work presented in  \cite{Chen:2021mtn}. We consider  $\cN=8$ gauged supergravity theories  which have a  $\cN=(4,4)$  $AdS_3$  vacuum  with  $D^1(2,1;\alpha) \times D^1(2,1;\alpha)$ symmetry for general $\alpha$. We derive the BPS flow equations and find numerical solutions. Some holographic quantities such as the entanglement entropy  are calculated.

\end{abstract}

\setcounter{equation}{0}
\setcounter{footnote}{0}

\newpage

\section{Introduction}

Janus solutions provide a holographic description of   interface conformal field theories.    Generally, the solutions are constructed by considering  an $\AdS_d$ slicing of a higher dimensional space where the other  fields depend non-trivially on the slicing coordinate(s).  For example  the  original Janus solution  \cite{Bak:2003jk},   deforms the $\AdS_5\times S^5$ vacuum of type IIB and is given by an $\AdS_4$ slicing  where the dilation depends non-trivially on a single slicing coordinate and approaches two different values on the two boundary components. The solution is dual to an interface of $N=4$ super Yang-Mills theory where the coupling $g_{\rm YM}$ jumps across a co-dimension one interface \cite{Clark:2004sb}.  More general  Janus solutions preserving  supersymmetry  were constructed  as  $\AdS_4 \times S^2 \times S^2$ space warped over a Riemann surface \cite{DHoker:2007zhm}. These solutions are dual to supersymmetric interface theories in $N=4$ SYM \cite{DHoker:2006qeo, Gaiotto:2008ak, Gaiotto:2008sd}. For other  Janus solutions in  ten and eleven dimensions, see e.g.~\cite{DHoker:2006vfr,DHoker:2009lky,DHoker:2007hhe,Bachas:2013vza}. In general, constructing  such solutions is quite difficult due to the fact that the supersymmetry variations, as well the equations of motion,  depend on more than one warping coordinate and the resulting equations are nonlinear partial differential equations.
A useful approach is to construct Janus solutions in lower dimensional gauged supergravities (see  for example  \cite{Pilch:2015dwa,Gutperle:2017nwo,Suh:2011xc,Bobev:2013yra,Clark:2005te,Suh:2018nmp,Karndumri:2016tpf,Karndumri:2020bkc,Chiodaroli:2009yw,Assawasowan:2022yni,Karndumri:2021pva}). Such solutions are often easier to obtain, can be uplifted to ten or eleven dimensions or can be used to explore qualitative features of Janus solution in  a bottom-up approach.

In lower dimensional gauged supergravities it is often the case that  in addition to a maximally supersymmetric AdS vacuum there are extrema with a reduced amount of supersymmetry. One of the aims of the present paper is to construct holographic Janus solutions which correspond to RG interfaces \cite{Gaiotto:2012np}, between different AdS vacua\footnote{See \cite{Gutperle:2012hy,Arav:2020asu,Korovin:2013gha} for other examples of holographic RG-flow interfaces.}. This paper is a continuation of the work presented  \cite{Chen:2021mtn}, which considered  three-dimensional $\cN=8$ gauged supergravity with $n=4$ vector multiplets,  first discussed in \cite{Nicolai:2001ac}.  This theory has an $\AdS_3$ vacuum with maximal $\cN=(4,4)$ supersymmetry as well as two families of $\AdS_3$ vacua with $\cN=(1,1)$ supersymmetry \cite{Berg:2001ty}. The gauged supergravity has  a parameter $\alpha$ on which the embedding tensor for the gauged supergravity depends. For this theory the dual    superconformal  algebra of the $\cN=(4,4)$ vacuum is given by the ``large'' superconformal algebra $D^1(2,1;\alpha) \times D^1(2,1;\alpha)$, and the three-dimensional supergravity is believed to be a truncation  of M-theory on $\AdS_3\times S^3\times S^3 \times S^1$ \cite{Boonstra:1998yu,Elitzur:1998mm,deBoer:1999gea,Gukov:2004ym}. In the previous paper we considered the special case of $\alpha=1$ for which the explicit expressions become simpler. Here we will analyze the case for general $\alpha$, using both analytical and numerical methods.

The structure of this  paper is as follows: In section \ref{sec2}  we review the three dimensional gauged supergravity with $n=4$  vector  multiplets used here. We consider three truncations where the gauge fields as well as some scalars can consistently  be set to zero  and fix the $\cN=(1,1)$ vacua for general $\alpha$. In section \ref{sec3} we derive the BPS flow equations for an $AdS_2$ sliced Janus ansatz, this generalizes and streamlines the discussion of \cite{Chen:2021mtn}. In section \ref{sec4} we present the flow equations  for the three truncations and integrate them numerically for the three truncations. For the second and third truncations where $\cN=(1,1)$ AdS vacua exists we present examples of RG-flow interfaces. In section \ref{sec5} we use the solutions to calculate some holographic observables. In particular we determine the masses of the fluctuating scalars around the $\cN=(1,1)$ vacua.  The mass squared  of the scalar fluctuations is  positive and quite large, which means that the scalar fluctuations around the  fixed point are repulsive in the UV. This implies that  the initial conditions have to be fine tuned in order to reach the fixed point.  We discuss our results and possible directions for future research in section \ref{sec6}.

%%%%%%%%%%%%%%%%
\section{Three-dimensional $\cN=8$ gauged supergravity}\label{sec2}
%%%%%%%%%%%%%%%%
In this section, we review the  $\cN=8$ gauged supergravity first constructed in  \cite{Nicolai:2001ac} mainly following the conventions of   \cite{Chen:2021mtn}.	The bosonic field content consists of a graviton $g_{\mu\nu}$, Chern-Simons gauge fields $B^\cM_\mu$, and scalars fields living  in a $G/H = \SO(8, n)/\SO(8) \times \SO(n)$ coset, which has $8n$ degrees of freedom before gauging.
The scalar fields are parametrized by a $G$-valued matrix $L(x)$ in the vector representation, which transforms under $H$ and the gauge group $G_0 \subseteq G$ by right and left multiplication of group elements  respectively.
	\begin{align}
	L(x) \to g_0(x)L(x)h^{-1}(x)
	\end{align}
for $g_0 \in G_0$ and $h \in H$.
The Lagrangian is invariant under such transformations.
In this paper we use the following index conventions:
	\begin{itemize}
	\item $I, J, \dotsc = 1, 2, \dotsc, 8$ for $\SO(8)$.
	\item $r, s, \dotsc = 9, 10, \dotsc, n+8$ for $\SO(n)$.
	\item $\bar I, \bar J,\dotsc = 1, 2, \dotsc, n+8$ for $\SO(8, n)$.
	\item $\cM, \cN, \dotsc$ for generators of $\SO(8, n)$. 	\end{itemize}	
Let the generators of $G$ be $\{t^\cM\} = \{t^{\bar I \bar J} \} = \{X^{IJ}, X^{rs}, Y^{Ir}\}$, where $Y^{Ir}$ are the noncompact generators.
Explicitly, the generators of the vector representation are given by
	\begin{align} 
	\tensor{(t^{\bar I \bar J})}{^{\bar K}_{\bar L}} = \eta^{\bar I \bar K} \delta^{\bar J}_{\bar L} - \eta^{\bar J \bar K} \delta^{\bar I}_{\bar L}
	\end{align}
where $\eta^{\bar I \bar J} = \diag( + + + + + + + + - \cdots)$ is an  $\SO(8,n)$-invariant tensor.
These generators satisfy the typical $\SO(8, n)$ commutation relations,
	\begin{align} 
	[t^{\bar I \bar J}, t^{\bar K \bar L}] = 2\qty( \eta^{\bar I [\bar K} t^{ \bar L ] \bar J} - \eta^{\bar J [\bar K} t^{ \bar L ] \bar I} )
	\end{align}

The gauging of the supergravity is characterized by an embedding tensor $\Theta_{\cM \cN}$ (which has to satisfy various identities \cite{deWit:2003ja} in order to define a consistent theory) that determines which isometries are gauged, the coupling to the Chern-Simons fields, and additional terms in the supersymmetry transformations and action depending on the gauge coupling $g$.
We will look at the particular case in \cite{Berg:2001ty} where $n \geq 4$ and the gauged subgroup is the $G_0 = \SO(4) \times \SO(4)$ subset of  the $\SO(8) \subset \SO(8, n)$.
The embedding tensor has the non vanishing  entries,\footnote{We use the conventions $\epsilon_{1234} = \epsilon_{5678} = 1$.}
	\begin{align} 
	\Theta_{\bar I \bar J, \bar K \bar L} = \begin{cases} \alpha \epsilon_{\bar I \bar J \bar K \bar L} & \text{if } \bar I, \bar J, \bar K, \bar L \in \{1, 2, 3, 4\} \\
		\epsilon_{\bar I \bar J \bar K \bar L} & \text{if } \bar I, \bar J, \bar K, \bar L \in \{5, 6, 7, 8\} \\ 0 & \text{otherwise} \end{cases}
	\end{align}
Note that the gauging depends on a real parameter $\alpha$.  
As discussed in  \cite{Berg:2001ty}, the maximally supersymmetric $\AdS_3$ vacuum has an isometry group,
	\begin{align}
	D^1(2,1;\alpha) \times D^1(2,1;\alpha)
	\end{align}
which corresponds to the family of ``large'' superconformal algebras of the dual SCFT.
In this paper we generalize the analysis of \cite{Chen:2021mtn} where the case $\alpha=1$ was considered to the case of general $\alpha$.  Note that in the special  case $\alpha=1$ the super algebra becomes more familiar  $D^1(2, 1; 1) = \OSp(4|2)$.

From the embedding tensor, the $G_0$-covariant currents can be obtained,
	\begin{align} 
	L^{-1} (\partial_\mu + g \Theta_{\cM \cN} B_\mu^\cM t^\cN ) L = \frac{1}{2} \cQ^{IJ}_\mu X^{IJ} + \frac{1}{2} \cQ^{rs}_\mu X^{rs} + \cP^{Ir}_\mu Y^{Ir}
	\end{align}
It is convenient to define the $\tensor{\cV}{^\cM_\cA}$ tensors,
	\begin{align}  
	L^{-1} t^\cM L = \tensor{\cV}{^\cM_\cA} t^\cA = \frac{1}{2} \tensor{\cV}{^\cM_{IJ}} X^{IJ} + \frac{1}{2} \tensor{\cV}{^\cM_{rs}} X^{rs} + \tensor{\cV}{^\cM_{Ir}} Y^{Ir}
	\end{align}
and the $T$-tensor,
	\begin{align} 
	T_{\cA | \cB} = \Theta_{\cM \cN} \tensor{\cV}{^\cM_\cA} \tensor{\cV}{^\cN_\cB}
	\end{align}
The $T$-tensor is used to  construct the  tensors $A_{1, 2, 3}$ which will appear in the scalar potential and the supersymmetry transformations,
	\begin{align} \label{amatrix}
	A_1^{AB} &= - \frac{1}{48} \Gamma^{IJKL}_{AB} T_{IJ|KL} \nonumber \\
	A_2^{A\dot A r} &= - \frac{1}{12} \Gamma^{IJK}_{A\dot A} T_{IJ|Kr} \nonumber \\
	A_3^{\dot A r \dot B s} &=  \frac{1}{48} \delta^{rs} \Gamma^{IJKL}_{\dot A \dot B} T_{IJ|KL} + \frac{1}{2} \Gamma^{IJ}_{\dot A \dot B} T_{IJ|rs}
	\end{align}
where $A, B$ and $\dot A, \dot B$ are $\SO(8)$-spinor indices.
Our conventions for the $\SO(8)$ Gamma matrices are presented in appendix \ref{appendix-gamma}.

Here  we choose the spacetime signature $\eta^{ab} = \diag(+--)$ as mostly negative.
The bosonic Lagrangian and scalar potential are given by 
	\begin{align} \label{lagrangian}
	e^{-1} \cL_{\rm bos} &= - \frac{1}{4} R + \frac{1}{4} \cP_\mu^{Ir} \cP^{Ir\, \mu} + V - \frac{1}{4} e^{-1} \epsilon^{\mu\nu\rho} g \Theta_{\cM \cN} B_\mu^\cM \qty( \partial_\nu B_\rho^\cN + \frac{1}{3} g \Theta_{\cK \cL} \tensor{f}{^{\cN \cK}_{\cP}} B_\nu^\cL B_\rho^\cP ) \nonumber \\
	V &= \frac{1}{4} g^2 \qty( A^{AB}_1 A^{AB}_1 - \frac{1}{2} A^{A \dot A r}_2 A^{A \dot A r}_2 ) 
	\end{align}
The supersymmetry variations take the following form
	\begin{align}
	\delta \chi^{\dot A r} &= \frac{1}{2} i \Gamma^I_{A\dot A} \gamma^\mu \epsilon^A \cP^{Ir}_\mu + g A^{A \dot A r}_2 \epsilon^A \nonumber \\
	\delta \psi^A_\mu &= \qty(\partial_\mu \epsilon^A + \frac{1}{4} \omega_\mu^{ab} \gamma_{ab} \epsilon^A + \frac{1}{4} \cQ^{IJ}_\mu \Gamma^{IJ}_{AB} \epsilon^B) + i g A^{AB}_1 \gamma_\mu \epsilon^B 
	\end{align}
The Einstein equations of motion are
	\begin{align} 
	R_{\mu\nu} - \cP^{Ir}_\mu \cP^{Ir}_\nu - 4 V g_{\mu\nu} = 0 
	\end{align}
and the gauge field equations of motion  are
	\begin{align} \label{gaugefieldeq}
	e \cP^{Ir\, \lambda} \Theta_{\cQ \cM} \tensor{\cV}{^\cM_{Ir}} = \epsilon^{\lambda \mu\nu} \qty( \Theta_{\cQ \cM} \partial_\mu B^\cM_\nu  + \frac{1}{6} g B^\cM_\mu B^\cK_\nu \qty( \Theta_{\cM \cN} \Theta_{\cK \cL} \tensor{f}{^{\cN \cL}_\cQ} + 2\Theta_{\cM \cN} \tensor{f}{^{\cL \cN}_\cK} \Theta_{\cL \cQ} ) )
	\end{align}
	 	
\subsection{The $n=4$ case}\label{sec21}
	
The smallest number of  matter multiplets where multiple supersymmetric vacua exist is $n = 4$.
The symmetries of the theory   are  a local $G_0 = \SO(4) \times \SO(4)$ and a global $\SO(n)$ with $ n=4$.
Consequently,  the scalar potential only depends on $8 \cdot 4 - 3 \cdot 6 = 14$ fields out of the original $32$.
Moreover,  a further  consistent truncation outlined in \cite{Berg:2001ty} is performed where the coset representative depends  only on eight of the fourteen scalars.
	\begin{align}
	L &= \mqty( \cos A & \sin A \cosh B & \sin A \sinh B \\ -\sin A & \cos A \cosh B & \cos A \sinh B \\ 0 & \sinh B & \cosh B) \nonumber \\
	A &= \diag(p_1, p_2, p_3, p_4)~, \qquad B = \diag(q_1, q_2, q_3 ,q_4)
	\end{align}
	We will not display the general form of the tensors $A_1$ and $A_2$ defined in (\ref{amatrix}) here.
The scalar potential has terms up to order $\alpha^2$.

	\begin{align}
	g^{-2} V &= {1\over 2}+{1\over 4}\sum_i x_i^2-{1\over 4} \sum_{i<j<k} x_i^2 x_j^2x_k^2 -{1\over 2}\prod_i  x_i^2 + \alpha\left(- \prod_i x_i y_i+\prod_i \sqrt{1+x_i^2+y_i^2}\right)\nonumber\\
	&+ \alpha^2 \left( {1\over 2}+{1\over 4}\sum_i y_i^2-{1\over 4} \sum_{i<j<k} y_i^2 y_j^2y_k^2 -{1\over 2}\prod_i  y_i^2 \right)
	\end{align}
	where all indices run form 1 to 4 unless otherwise indicated and we used the following definition of  scalar fields
	\begin{align}\label{xyqp}
x_i &= \cos p_i \sinh q_i ~, \qquad y_i = \sin p_i \sinh q_i
\end{align}
The $\cQ_\mu$ and $\cP_\mu$ currents do not depend on $\alpha$, excluding the $g \Theta_{\cM\cN} B^\cM_\mu \tensor{\cV}{^\cN_\cA}$ term, they are given by
	\begin{align}
	\cQ^{IJ}_\mu &=  \qty(\smqty{ 0&0&0&0& \cosh q_1 \partial_\mu p_1 &0&0&0 \\
						0&0&0&0& 0& \cosh q_2 \partial_\mu p_2 &0&0 \\
						0&0&0&0& 0&0& \cosh q_3 \partial_\mu p_3 &0 \\
						0&0&0&0& 0&0&0& \cosh q_4 \partial _\mu p_4 \\
						-\cosh q_1 \partial_\mu p_1 &0&0&0 & 0&0&0&0 \\
						0& -\cosh q_2 \partial_\mu p_2 &0&0 & 0&0&0&0 \\
						0&0& -\cosh q_3 \partial_\mu p_3 &0 & 0&0&0&0 \\
						0&0&0& -\cosh q_4 \partial _\mu p_4 & 0&0&0&0 })_{IJ}
	\nonumber \\
	\cQ^{rs}_\mu &= 0 \nonumber \\
	\cP^{Ir}_\mu &= \qty(\smqty{ \sinh q_1 \partial_\mu p_1 & 0 & 0 & 0 \\
						  0 & \sinh q_2 \partial_\mu p_2 & 0 & 0 \\
						  0 & 0 & \sinh q_3 \partial_\mu p_3 & 0 \\
						  0 & 0 & 0 & \sinh q_4 \partial_\mu p_4 \\
						  \partial_\mu q_1 & 0 & 0 & 0 \\
						  0 & \partial_\mu q_2 & 0 & 0 \\
						  0 & 0 & \partial_\mu q_3 & 0 \\
						  0 & 0 & 0 & \partial_\mu q_4 })_{Ir}
	\end{align}
Using these matrices, we can check that the combination $\cP^{Ir}_\mu \tensor{\cV}{^{JK}_{Ir}} $ vanishes whenever the indices $J, K \in \{1, 2, 3, 4\}$ or $J, K \in \{5, 6, 7, 8\}$.
This implies that there is no source for $B^\cM_\mu$ in the gauge field equation of motion (\ref{gaugefieldeq}), so it is consistent to set $B^\cM_\mu = 0$.
We will make this choice from now on.

The kinetic term for the scalars in the action (\ref{lagrangian}) can be expressed in terms of the $x_i$ and $y_i$ using the relations (\ref{xyqp}) and takes the form
\begin{align}
 \frac{1}{4} \cP_\mu^{Ir} \cP^{Ir\, \mu} = -{1\over 4} \sum_{i=1}^4 {1\over 1+ x_i^2+y_i^2} \Big( (1+ y_i^2) (\partial_\mu \partial^\mu x_i - 2 x_i y_i \partial_\mu  x_i \partial^\mu y_i' + (1+ x_i^2) \partial y_i\partial^\mu y_i \Big)
\end{align}
This expression will be needed for determining masses of the fluctuations of the scalar fields around the supersymmetric vacua.
	
\subsection{Truncations and supersymmetric $\AdS_3$ vacua}
\label{sec:trunc3}
	
In order to make our analysis more tractable, we make further truncations to reduce the number of independent scalar fields. Below we consider three consistent  truncations, which together explore the  $\AdS_3$ vacua with $\cN = (4,4)$ and $\cN = (1,1)$ supersymmetry. All of the results  are generalizations of the $\alpha=1$ case discussed in  \cite{Chen:2021mtn}.

\subsubsection{Truncation 1}\label{trunc1}

The first truncation is given by denoting  $q_1 = q$, $p_1 = p$ and setting all remaining $q_i = p_i = 0$ for $i = 2,3,4$. 
The scalar potential is
	\begin{align}
	V = \frac{g^2}{4}  \left( 2(1+\alpha^2) + 4 \alpha \cosh q+  (\cos^2 p+\alpha^2 \sin^2 p)\sinh^2 q \right)
	\end{align}
The $\cN=(4,4)$ vacuum is given  by setting $q=0$ and the vacuum potential is $V_0 = {1\over 2} g^2 (1+\alpha)^2$. In the $x,y$ coordinates the $\cN=(4,4)$ vacuum is given by $x_i=y_i=0$. This is the only supersymmetric vacuum for this truncation. We note that for the choice $\alpha=1$ the potential is independent of  the scalar  field $p$. We note we will chose $g= 1/(1+\alpha)$ in order to set the potential at the $\cN=(4,4)$ vacuum to be $V_0= {1\over 2}$,  which corresponds to a unit radius $AdS_3$.

\subsubsection{Truncation 2}\label{trunc2}

\begin{figure}  \centering
  \includegraphics[width=70mm]{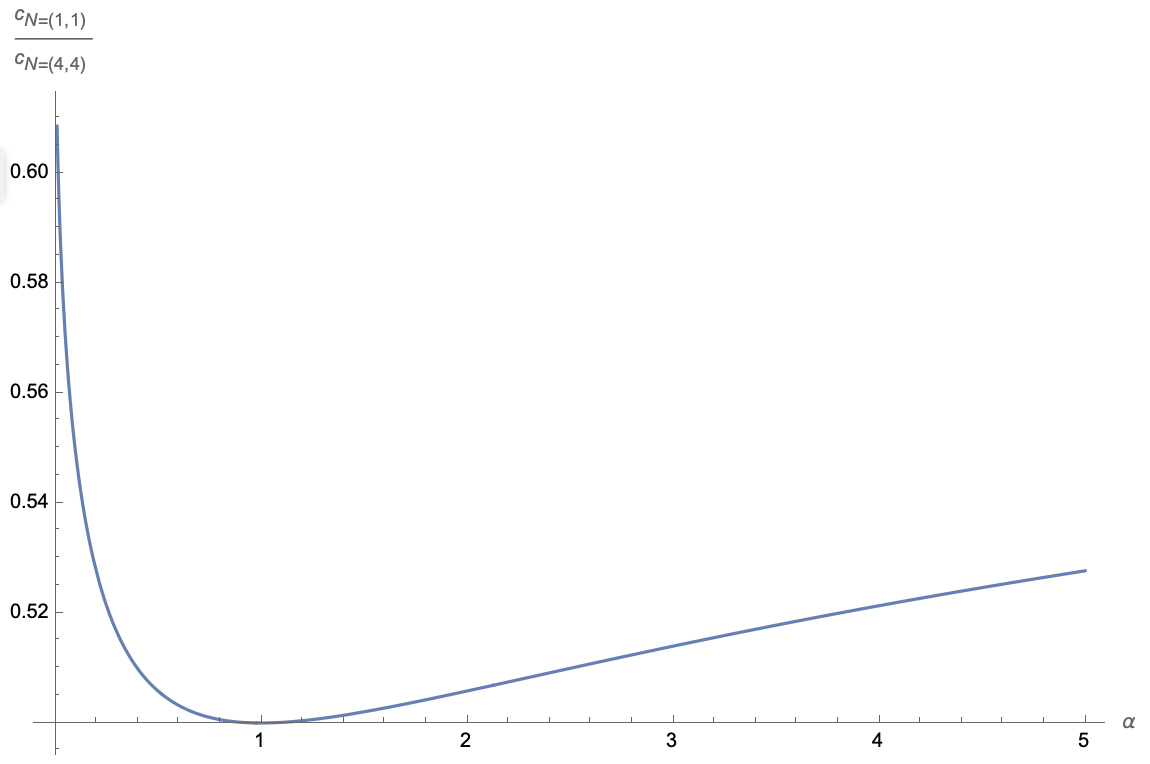}
  \caption{Ratios of central charges for the $\cN=(1,1)$ and $\cN=(4,4)$ vacua.}\label{fig:one}
\end{figure}

The second truncation is given by setting all the $q$ and $p$ equal, i.e.~$q_i=q$, $p_i=p$ for $i=1,2,3,4$.  
The scalar potential becomes
	\begin{align}
	V &= {g^2\over 2} \Big\{ (1- \cos^2 p \sinh^2q) (1+  \cos^2 p \sinh^2q)^3+\alpha^2 (1- \sin^2 p \sinh^2q) (1+  \sin^2 p \sinh^2q)^3\nonumber \\
	&\quad +\alpha ( 2+ 4 \sinh^2 q+ 2 \sinh^4q- 2 \sin^4 p \cos^4 p \sinh ^8q)\Big\}
		\end{align}
or in terms of the $x,y$ fields, the potential will take the following form
		\begin{align}
		V&= {g^2\over 2} \Big\{ (x^2-1) (x^2+1)^3 +\alpha^2 (y^2-1)(y^2+1)^3 + 2 \alpha \big( 2x^2(1+y^2) +(1+y^2)^2- x^4(y^4-1)\big) \Big\}
		\end{align}
As before the $\cN=(4,4)$ vacuum is given by $q=0$ or $x=y=0$. There are $\cN=(1,1)$ vacua   which are located at
\begin{align} \label{xy2trun}
x&= \pm {1\over 6 \sqrt{3} \alpha} \left(-12 \alpha^2- {2^{2\over 3} Y^{1\over 3}(Y+ 2\alpha^2(-3 + 8 \alpha))\over  (3+2 \alpha)} -{2^{1\over 3} y^{2\over 3}(Y+ 2\alpha(-18 + \alpha (-15+8\alpha))) \over (3+ 2\alpha)^2}\right)^{1\over 2}\nonumber\\
y&= \pm {1\over 3} \Big( -1 + {2^{2\over 3} Y^{1\over 3} \over \alpha} + {4 \; 2^{1\over 3} (3+ 2\alpha)\over Y^{1\over 3}}\Big)^{1\over 2}
\end{align} 
where $Y$ is given by
\begin{align}
Y= 3i \alpha^{3\over 2}  \sqrt{96 + 3 \alpha(61+32 \alpha)}-\alpha^2(9+16 \alpha)
\end{align}
The central charge of the dual CFT is related to the AdS radius and the value of the potential $V_0$ at its minimum
\begin{align}
c&={3 R_{AdS}\over 2 G_N} = {3\over  \sqrt{2 V_0} } {1\over 2 G_N}
\end{align}
Choosing $g=1/(1+\alpha)$ sets the AdS radius of the $\cN=(4,4)$ vacuum to one and the ratio of the central charge of the $\cN=(4,4)$ to the $\cN=(1,1)$ vacuum as a function of $\alpha$ becomes 
\begin{align}\label{ratiocs}
{c_{\cN=(1,1)} \over c_{\cN=(4,4)} } = {1\over \sqrt{ 2 V_0^{N=(1,1)}(\alpha)}}
\end{align}
The expressions derived in (\ref{xy2trun}) are not very illuminating and we present a plot of the ratio of the central charges for the two vacua in the  figure \ref{fig:one}. It is interesting to note that the ratio of central charges is minimized  for the special value $\alpha=1$.

\subsubsection{Truncation 3}\label{trunc3}

The third truncation is given by setting the first three $q$ and $p$ equal, i.e.~$q_i = q$, $p_i = p$ for $i = 1, 2, 3$, and setting the remaining $q_4 = p_4 = 0$.
The scalar potential is 
	\begin{align}
	V&={g^2\over 4} \Big( 2+ 3 \cos^2 p \sinh^2 q- \cos^6 p \sinh^6q + 4 \alpha \cosh^3 q + \alpha^2 ( 2+ 3 \sin^2 p \sinh^2 q-\sin^6 p \sinh^6q)\Big)
	\end{align}
	or in the $x,y$ variables
	\begin{align}\label{vxytrun3}
	V&={g^2\over 4} \Big(  (2+ 3 x^2-x^6) + 4  \alpha (1+ x^2+y^2)^{3\over 2} + \alpha^2(2+ 3y^2-y^6)\Big)
	\end{align}
The $\cN=(4,4)$ vacuum is given by $q=0$  or $x=y=0$ as before, and $\cN=(1,1)$ vacua can be determined  by finding the extrema for the potential (\ref{vxytrun3}) away from the origin.  
\begin{align}
y&= \pm \sqrt{{1\over 2} }\left( 1+\ep(\alpha)  \sqrt{1- {4\over X^{1\over 3}}+ {2 X^{1\over 3} \over 3 \alpha^2}}+\sqrt{ 2+ {4\over X^{1\over 3}} -{2 X^{1\over 3}\over 3 \alpha^2}+ {2(\alpha^2-4) \over \alpha^2 \sqrt{1- {4\over X^{1\over 3}}+ {2 X^{1\over 3} \over 3 \alpha^2}}}}\right)^{1\over 2} \nonumber \\
x&= \pm \left( {\alpha^2\over 4} (y^4-1)^4 -1-y^2\right)^{1\over 2} 
\end{align}
where we used the abbreviation
\begin{align}
X= 3 \alpha^2 \big(9-9\alpha^2 +\sqrt{81-138 \alpha^2+81 \alpha^4}\big)
\end{align}
The $\epsilon(\alpha)$ is a sign which selects a branch of the solutions which gives real $x,y$ depending on $\alpha$ and we have $\epsilon(\alpha)=+1$ for $\alpha<2$ and $\epsilon(\alpha)=-1$ for $\alpha>2$.  We can plot the ratio of the central charges which is given by (\ref{ratiocs}), determined from the potential (\ref{vxytrun3}). 
\begin{figure}  \centering
  \includegraphics[width=70mm]{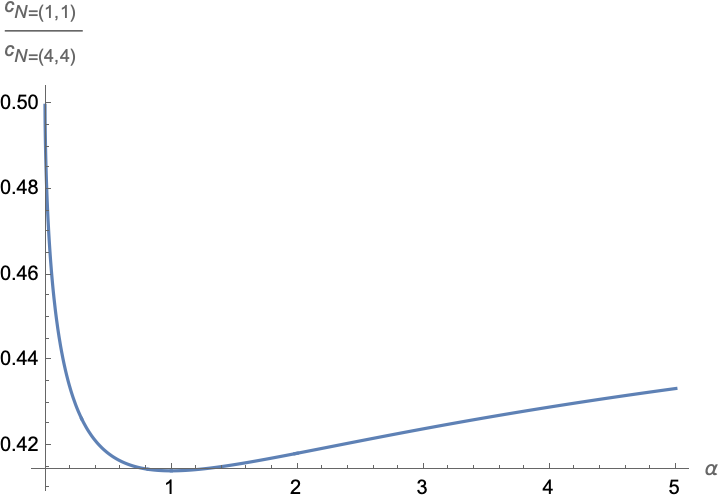}
  \caption{Ration of central charges for the $\cN=(1,1)$ and $\cN=(4,4)$ vacua.}\label{fig:two}
\end{figure}
We note that the qualitative behavior of the ratio for truncation 2 and 3 is very similar, in particular the central charge for the $\cN=(1,1)$ vacuum is minimized at $\alpha=1$.

%%%%%%%%%%%%%%%%
\section{Janus flow equations }\label{sec3}
%%%%%%%%%%%%%%%%
In this section we will  derive  the BPS flow equations, expanding on the construction in our previous paper   \cite{Chen:2021mtn}. The Janus ansatz for the bosonic fields is give  by
\begin{align} \label{janusan}
	\dd{s^2} &= e^{2B(u)} \qty(\frac{ \dd{t^2} - \dd{z^2} }{z^2}) -\dd{u}^2~,  \qquad
	q_i = q_i(u)~, \qquad  p_i = p_i(u)
	\end{align}
The Chern-Simons gauge fields is set  to zero $ B^\cM_\mu  = 0$. We will  check that the source term on the right hand side of the gauge field equation of motion (\ref{gaugefieldeq}) is zero for the solutions considered in this paper.

	The gravitino supersymmetry  variation  $\delta \psi^A_\mu = 0$ is
	\begin{align}\label{gravvar}
	0 &= \partial_t \epsilon + \frac{1}{2z} \gamma_0 \qty(\gamma_1  - B' e^{B} \gamma_2 + 2 i g e^B A_1) \epsilon \nonumber \\
	0 &= \partial_z \epsilon + \frac{1}{2z} \gamma_1 \qty( - B' e^{B} \gamma_2 + 2 i g e^B A_1 ) \epsilon \nonumber \\
	0 &= \partial_u \epsilon + \frac{1}{4} \cQ_u^{IJ} \Gamma^{IJ} \epsilon + i g  \gamma_2 A_1 \epsilon 
	\end{align}
where we have suppressed the $\SO(8)$-spinor indices of $\epsilon^A$ and $A_1^{AB}$. 
The spin-${1\over 2}$  variation $\smash{\delta \chi^{\dot A r} = 0}$ is
	\begin{align} \label{integrac}
	\qty(- \frac{i}{2} \Gamma^I \cP_u^{I r} \gamma_2 + g A_2^{r} )_{A\dot A} \epsilon^A = 0~, \quad \quad r=9,10,\dotsc, 8+n
	\end{align}
The matrix $A_1$ defined in (\ref{amatrix})  has eigenvectors
	\begin{align}
	A_1^{AB} n_\pm^{(i) B} = \pm w_i   n_{\pm}^{(i) A}~, \quad \quad i=1,2,3,4
	\end{align}
	For a supersymmetric $AdS_3$ vacuum the eigenvalue $w_i$ is related to the value of the potential evaluated at the vacuum via
	\begin{align}\label{susyvac}
	w^2_{\rm vac} = {V_{\rm vac}\over 2 g^2} 
	\end{align}
and the associated eigenvectors $ n_{\pm}^{(i)}$ determine the supersymmetries of the vacuum.
	For the $\cN=(4,4)$ vacuum the $w_i, i=1,\cdots 4$ all satisfy  (\ref{susyvac}) and hence the vacuum preserves eight supersymmetries.  For the $\cN=(1,1)$ vacuum only one of the four $n_{\pm}^{(i)}$ and $w_i$ satisfies  (\ref{susyvac}). In the following we drop the index $^{(i)}$ to denote the supersymmetric eigenvalue $w$ and the eigenvector $n_\pm^A$.

The general ansatz for unbroken supersymmetry   $\epsilon^A$   for the Janus solution is given by
	\begin{align} \label{spinor-ansatz}
	\epsilon^A  =  \big(f_+  n^A_+  + f_- n_-^{ A}\big) \zeta_{+}+\big(g_+ n_+^{A} + g_-^{} n_-^{ A}\big) \zeta_{-}
	\end{align}
where $\zeta_\pm$ are Killing spinors for a unit radius $AdS_2$ 
\begin{align}
	D_\mu \zeta_\eta  = i {\eta\over 2}\gamma_\mu \zeta_\eta~, \quad \quad \mu=t, z, \quad \eta =\pm1
	\end{align} 

\subsection{Gravitino variation}
The $t, z$ components of the gravitino variation 
can be expressed as follows by using the properties of the $\AdS_2$ Killing spinors,
	\begin{align}
	0&= i \qty{ \big(f_+  n_+^{   A} + f_-^{ } n_-^{  A}\big) \zeta_{+} - \big(g_+^{ } n_+^{  A} + g_-^{ } n_-^{  A}\big)  \zeta_{-} } \\
	&\quad  + i B'e^{B} i \gamma_2 \Big\{ \big(f_+^{ } n_+^{  A} + f_-^{ } n_-^{  A}\big) \zeta_{+}+\big(g_+^{ } n_+^{  A} + g_-^{ } n_-^{  A}\big) \zeta_{-}\Big\}\nonumber\\
	&\quad + 2i g w e^{B} \Big\{ \big(f_+^{ } n_+^{  A} - f_-^{ } n_-^{  A}\big) \zeta_{+}+\big(g_+^{ } n_+^{  A} - g_-^{ } n_-^{  A}\big) \zeta_{-}\Big\}
	\end{align}
Using $i\gamma_2 \zeta_{\eta }= \zeta_{-\eta}$ and the linear independence of the $n_{\pm} $  and $\zeta_{\pm}$, one obtains a set of equations,
	\begin{align}\label{fgeq}
	f_+ + B' e^B g_+ + 2 g w e^B f_+&=0\nonumber \\
	-g_+ + B' e^B f_+ + 2 g w e^B g_+&=0\nonumber \\
	f_- + B' e^B g_- - 2 g w e^B f_-&=0\nonumber \\
	-g_- + B' e^B f_- - 2 g w e^B g_-&=0
	\end{align}
	It is convenient  to define the following expressions 
	\begin{align}\label{gamdef}
	\gamma(u) = \sqrt{1- {e^{-2B}\over 4 g^2 w^2}}~, \quad \quad \sqrt{1-\gamma^2(u)}= {e^{-B}\over 2g w}
	\end{align}
The equations (\ref{fgeq}) can then   be  solved by 
	\begin{align}\label{freplace}
	f_+ = {\sqrt{1-\gamma^2} -1\over \gamma}g_+~, \qquad  f_- = {\sqrt{1-\gamma^2} +1\over \gamma}g_-
	\end{align}
 if the integrability condition
\begin{align}\label{bprime}
	B'&= \pm \sqrt{4 g^2 w^2 -e^{-2B}} \nonumber\\
	&= \pm 2 g w \gamma
	\end{align}
is satisfied. This equation provides us with a differential equation for the metric factor $B$.  

\subsection{Spin ${1\over 2}$ variation}

The spin-${1\over 2}$  variation  (\ref{integrac})  takes the following form of a projector
	\begin{align}\label{proj1}
	\Big(M^{(r)AB} i \gamma_2 + \delta^{AB} \Big) \epsilon^B = 0
	\end{align}
	where
	\begin{align} \label{mmatrix-flow}
	M^{(r)}_{AB}= - \frac{1}{2g} \Big( \Gamma^I \cP_u^{I r} (A_2^{r})^{-1}  \Big)^T _{AB} 
	\end{align}
	Note that there is a projector for each $r$, which all have to be satisfied and the resulting flow equations are mutually consistent for a supersymmetric Janus solution to exist. This analysis will be performed for the particular truncations presented in section \ref{sec:trunc3}.
	
Inserting $\epsilon^A$  given by  (\ref{spinor-ansatz})   into the spin ${1\over 2}$ projector gives
\begin{align} \label{spinhalfeq}
	0&= \big(f_+^{ } n_+^{  A} + f_-^{ } n_-^{  A}\big) \zeta_{+}+\big(g_+^{ } n_+^{  A} + g_-^{ } n_-^{  A}\big) \zeta_{-}\nonumber \\
	&\quad + M^{AB} i\gamma_2  \Big\{  \big(f_+^{ } n_+^{  B} + f_-^{ } n_-^{  B}\big) \zeta_{+}+\big(g_+^{ } n_+^{  B} + g_-^{ } n_-^{  B}\big)\zeta_{-}\Big\}
	\end{align}
	We have dropped the index $r$ for notational convenience. Using the fact that  the two dimensional Killing spinors are orthogonal we can project (\ref{spinhalfeq})
onto the $n_{\pm}^{ }$ and $\zeta_\pm$ components. This produces  four equations
	\begin{align}\label{fgcon}
	f_+  n_+^2 + M_{++} g_+ + M_{+-} g_-&=0\nonumber \\
	g_+  n_+^2  + M_{++} f_+ + M_{+-} f_-&=0\nonumber \\
	f_- n_-^2 + M_{+-} g_+ + M_{--} g_-&=0\nonumber \\
	g_+n_-^2  + M_{+-} f_+ + M_{--} f_-&=0 
	\end{align}
where we denoted $n_\pm^2= n_\pm^A n_\pm^A$ and  we define 
	\begin{align}
	M_{++}= n_{+}^A   M^{AB} n_+^B~, \qquad  M_{--}= n_{-}^A   M^{AB} n_-^B~, \qquad  M_{+-}= M_{-+}= n_{+}^A   M^{AB} n_-^B
	\end{align}
If there is more than one $n_\pm$ (as in truncation 1) one has to  choose linear combinations for which  $M_{\pm\pm}, M_{\pm\mp}$  take the same form for all $n^{(i)}_\pm$, which is a consistency condition. 
Using (\ref{freplace}) it can be shown that  equations  (\ref{fgcon}) can only  be satisfied if we have
	\begin{align}\label{mmatrix}
	M_{++}= \gamma n_+^2 ~, \qquad M_{--} =-\gamma n_-^2 ~, \qquad M_{+-}=M_{-+}=\sqrt{1-\gamma^2} \sqrt{n_+^2 n_-^2}
	\end{align}
	In all cases we consider, the $M_{--}$ equation is automatically satisfied if  the $M_{++}$ equation is satisfied.  Hence (\ref{mmatrix}) provides two independent equations. It follows from (\ref{mmatrix-flow}) that these equations are linear in the first derivatives of the scalar fields and provide the BPS flow equations  for the scalars. The complete set of flow equations is given by these equations and  the flow equation for the metric factor (\ref{bprime}), coming from the gravitino variation.

%%%%%%%%%%%%%%%%
\section{Janus and RG-flow solutions }\label{sec4}
%%%%%%%%%%%%%%%%

In this section we obtain the flow equations and solve them numerically for the three truncations considered  in this paper. Since the first truncation does not have $\cN=(1,1)$ vacua the BPS flows will correspond to Janus solutions interpolating between $\cN=(4,4)$ vacua.  For the two other truncations we find Janus as well as RG-flow interface solutions. 

\subsection{Truncation 1} \label{sec4-1}	
The matrix $A_1$ for this truncation is given by
\begin{align}
A_1= 
\left(
\begin{array}{cccccccc}
  0&0   &0  &a&0&0&b&0  \\
  0&0   &-a  &0&0&0&0&b  \\
0 &-a   &0  &0&-b&0&0& 0 \\
 a & 0  &0  &0&0&-b&0&0  \\
 0 &0   &-b  &0&0&0&0&-a  \\
 0 & 0  &0  &-b&0&0&a&0  \\
b  &0   &0  &0&0&a&0&0  \\
0  &b   &0  &0&-a&0&0&0  \\
\end{array}
\right) 
\end{align}
where 
\begin{align}\label{abdef}
a={1\over 2} \cos p (\alpha +\cosh q), \quad b={1\over 2} \sin p(1+\alpha \cosh q)
\end{align}
The eigenvalue of $A_1$ are $\pm w_0$ which is given by
\begin{align}\label{w0def}
w_0 = \sqrt{a^2+b^2}={1\over 2}\sqrt{ \cos^2 p (\alpha +\cosh q)^2+ \sin^2 p(1+\alpha \cosh q)^2}
\end{align}
The eigenvectors are given by
\begin{align}
n^{(1)}_\pm  =
\left(
\begin{array}{c}
0   \\
    {a\pm w_0\over b} \\
 {-a\mp w_0 \over b}\\
0 \\
 1\\
 0\\
 0\\
 1
\end{array}
\right), \quad n^{(2)}_\pm  
=
\left(
\begin{array}{c}
0   \\
    {-a\pm w_0\over b} \\
 {-a\pm w_0\over b}\\
0 \\
- 1\\
 0\\
 0\\
 1
\end{array}
\right), \quad  n^{(3)}_\pm  =
\left(
\begin{array}{c}
 {-a\pm w_0\over b}  \\
  0 \\
0 \\
{a\mp w_0\over b}\\
 0\\
 1\\
 1\\
 0
\end{array}
\right), \quad, n^{(4)}_\pm  =
\left(
\begin{array}{c}
{a\pm w_0\over b}\\
  0 \\
0 \\
 {a\pm w_0\over b}\\
 0\\
 -1\\
 1\\
 0
\end{array}
\right),
\end{align}
The matrix $M^{AB}$ defined in (\ref{mmatrix-flow}) takes the following form for the truncation 1
\begin{align}
M&= 
\left(
\begin{array}{cccccccc}
  0&0   &0  &m_1&0&0&-m_2&0  \\
  0&0   &-m_1  &0&0&0&0&-m_2 \\ 
  0&-m_1   &0  &0&m_2&0&0&0  \\
    m_1&0   &0  &0&0&0&m_2&0  \\
  0&0   &m_2  &0&0&0&0&-m_1  \\ 
  0&0   &0  &m_2&0&0&m_1&0  \\
 -m_2&0   &0  &0&0&m_1&0&0  \\ 
  0&-m_2   &0  &0&-m_1&0&0&0  \\
\end{array}
\right) 
\end{align}
 with
 \begin{align}
 m_1&= {\alpha \sin p\;  p'+ \cos p \csch q \; q'\over g(\cos^2p+\alpha^2 \sin^2p)}, \quad \quad m_2= {\cos p\; p'-\alpha \sin p \csch q \; q'\over  g(\cos^2p+\alpha^2 \sin^2p)},
 \end{align}
 
 \begin{figure}
\centering
\begin{subfigure}[b]{0.3\textwidth}
	\includegraphics[width=\textwidth]{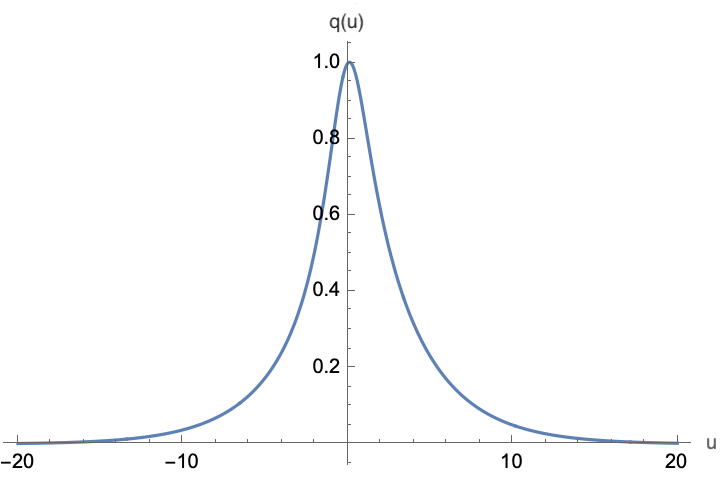}
	\vspace{-0.75cm}
	\caption{}
	\label{fig3-a}
\end{subfigure}
\hspace{1in}
\begin{subfigure}[b]{0.3\textwidth}
	\includegraphics[width=\textwidth]{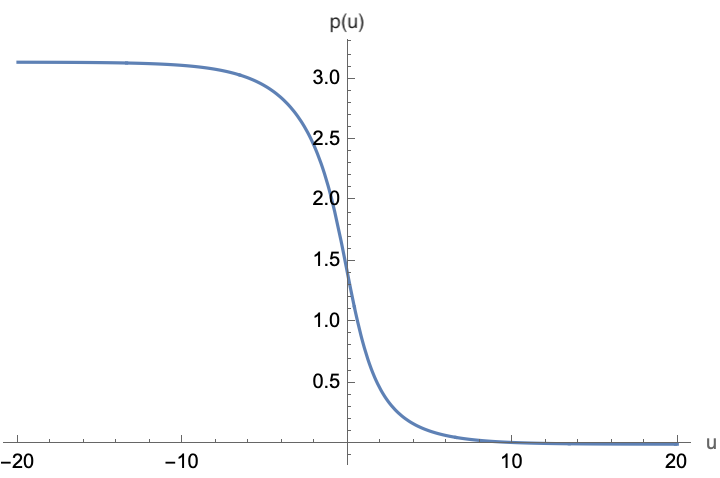}
	\vspace{-0.75cm}
	\caption{}
	\label{fig3-b}
\end{subfigure}
\\
\vspace{0.5cm}
\begin{subfigure}[b]{0.3\textwidth}
	\includegraphics[width=\textwidth]{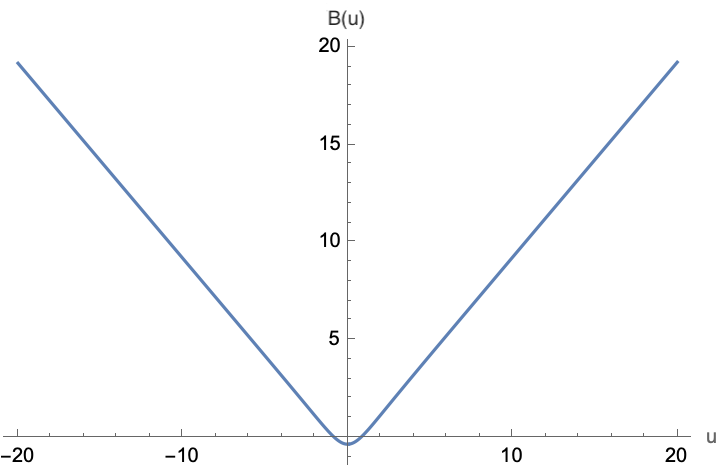}
	\vspace{-0.75cm}
	\caption{}
	\label{fig3-c}
\end{subfigure}
\hspace{1in}
\begin{subfigure}[b]{0.3\textwidth}
	\includegraphics[width=\textwidth]{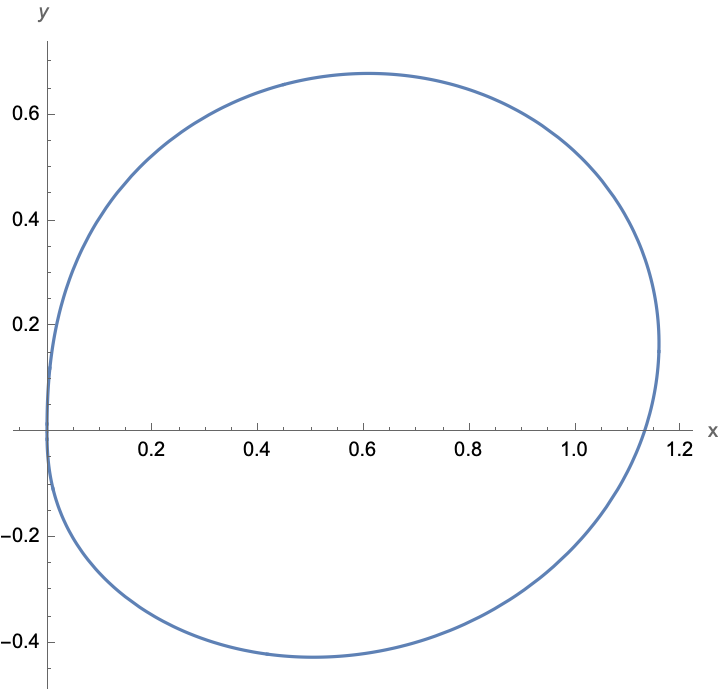}
	\vspace{-0.75cm}
	\caption{}
	\label{fig3-d}
\end{subfigure}
\caption{(a)-(c) plots of $q,p,B$ respectively, (d) parametric plot of the Janus flow in the $x,y$ variables. The initial conditions are $q(0)=1.0$ and $p(0) =1.5$ and $\alpha=2.3$.} \label{fig-trun1}
\end{figure}

 Using the definitions (\ref{abdef}) and (\ref{w0def}) the flow equations (\ref{mmatrix}) for the scalars $p,q$  and the metric function $B$ (\ref{bprime})  can be written relatively compactly
 \begin{align}\label{bpstrun1}
 p'&={g\over w_0} \Big( \alpha(a \gamma- b \sqrt{1-\gamma^2} )\sin p  -( b \gamma +a \sqrt{1-\gamma^2} )\cos p\Big)\nonumber\\
 q'&= {g\sinh q\over w_0} \Big( \alpha(b\gamma+a\sqrt{1-\gamma^2} )\sin p+ (a\gamma-b\sqrt{1-\gamma^2} )\cos p
\Big)\nonumber\\
B'&= \pm \sqrt {4 g^2w_0^2  - e^{-2B}}  \end{align}
This system of ordinary differential equations can only be integrated numerically. We will choose the coordinate $u$ such that the turning point of the metric function where $B'(u)=0$  is located at $u=0$. We then use the BPS equations  (\ref{bpstrun1}) to determine $p'(0)$, $q'(0)$ and $B(0)$  for a given $q(0)$ and $p(0)$. We then integrate the equations of motion following from the variation of the Lagrangian (\ref{lagrangian}). This means that all our solutions depend on two initial conditions $q(0)$ and $p(0)$. We have given an illustrative example  of the flows we can obtain in figure \ref{fig-trun1}.

\subsection{Truncation 2}\label{sec4:trun2}

The matrix $A_1$ for this truncation is given by
\begin{align}
A_1= 
\left(
\begin{array}{cccccccc}
  0&0   &0  &{a+c\over 4}&0&0&0&0  \\
  0&-{c\over 2}    &{-a+c\over 4} &0&b&0&0&b  \\
0 &{-a+c\over 4}   &-{c\over 2}  &0&-b&0&0& -b\\
 {a+c\over 4} & 0  &0  &0&0&0&0&0  \\
 0 &b   &-b  &0&{c\over 2} &0&0&{-a+c\over 4}   \\
 0 & 0  &0  &0&0&0&{a+c\over 4}&0  \\
0  &0   &0  &0&0&{a+c\over 4}&0&0  \\
0  &b  & -b &0&{-a+c\over 4} &0&0&{c\over2}  \\
\end{array}
\right) 
\end{align}
where 
\begin{align}
a&=2 \cos^4 p (\alpha \cosh^4 q)+ 2 \sin^4 p (1+ \alpha \cosh^4 q)\nonumber \\
b&= \sin p \cos p \cosh q \big( \cos^2 p (\alpha+\cosh^2 q) -\sin^2 p (1+\alpha \cosh^2q)\big)\nonumber \\
c&=(1+\alpha) \cosh^2 q \sin^2 2p
\end{align}
The eigenvectors $n^{(1)}_\pm$ of $A_1$  with eigenvalues $\pm w_0$ corresponding to the unbroken  $\cN=(1,1)$ supersymmetries  are given by
\begin{align}
n^{(1)}_\pm  =
\left(
\begin{array}{c}
0   \\
 {a-3c \pm 4 w_0 \over 8b} \\
 - {a-3c \pm 4 w_0 \over 8b}\\
0 \\
 1\\
 0\\
 0\\
 1
\end{array}
\right), \quad w_0= {1\over 4} \sqrt{ 64 b^2 + (3c-a)^2}
\end{align}

 We have checked that the extremum (\ref{xy2trun}) does satisfy the supersymmetry condition (\ref{susyvac})
for the $w_0$ defined above and hence corresponds to an AdS vacuum with $\cN=(1,1)$ supersymmetry. The rest of the eigenvectors of $A_1$ do not have eigenvalues which satisfy the supersymmetry condition (\ref{susyvac}) for the $\cN=(1,1)$ vacuum. We chose  the initial conditions the same way as in section \ref{sec4-1}. 
 \begin{figure}
\centering
\begin{subfigure}[b]{0.3\textwidth}
	\includegraphics[width=\textwidth]{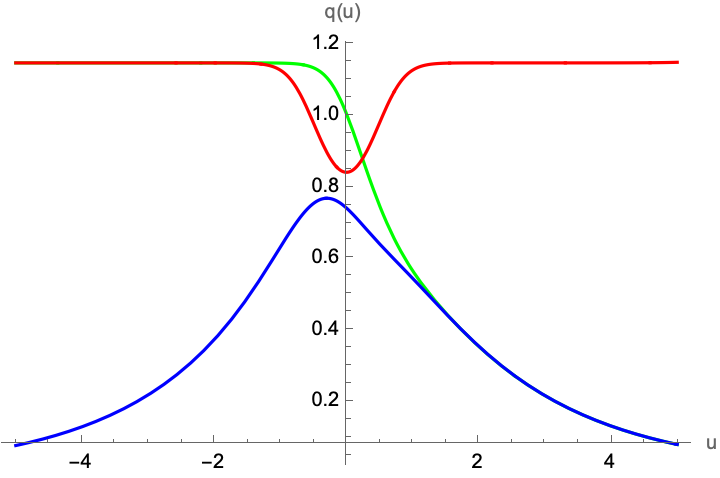}
	\vspace{-0.75cm}
	\caption{}
	\label{fig4-a}
\end{subfigure}
\hspace{1in}
\begin{subfigure}[b]{0.3\textwidth}
	\includegraphics[width=\textwidth]{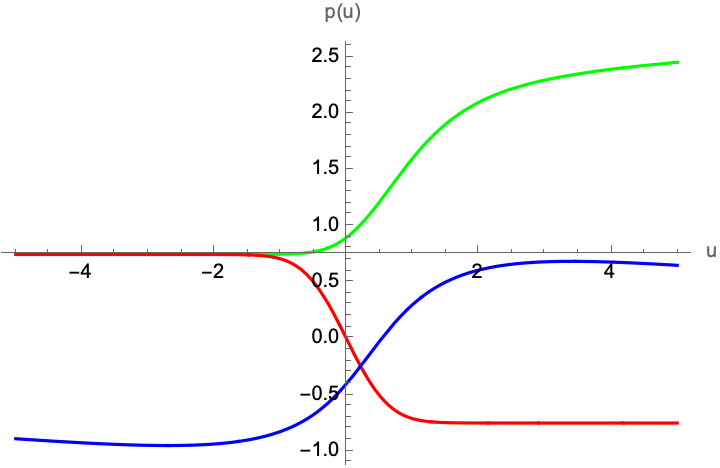}
	\vspace{-0.75cm}
	\caption{}
	\label{fig4-b}
\end{subfigure}
\\
\vspace{0.5cm}
\begin{subfigure}[b]{0.3\textwidth}
	\includegraphics[width=\textwidth]{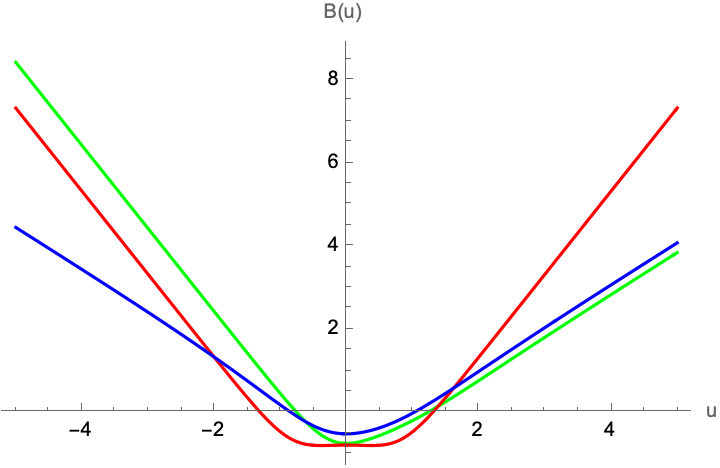}
	\vspace{-0.75cm}
	\caption{}
	\label{fig4-c}
\end{subfigure}
\hspace{1in}
\begin{subfigure}[b]{0.3\textwidth}
	\includegraphics[width=\textwidth]{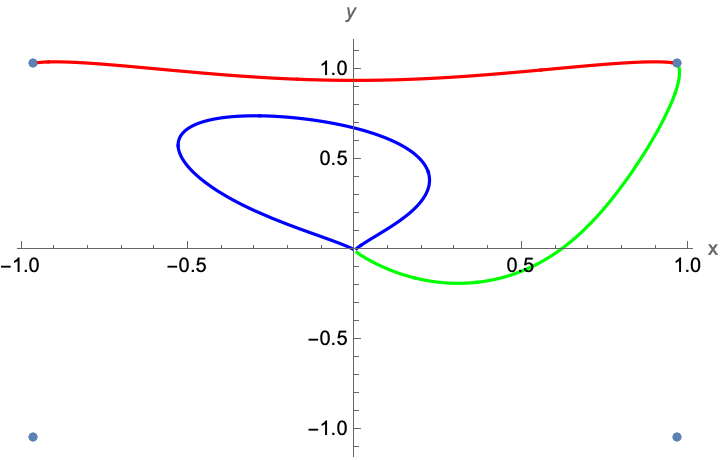}
	\vspace{-0.75cm}
	\caption{}
	\label{fig4-d}
\end{subfigure}
\caption{Truncation 2: (a)-(c) plots of $p,q,B$ respectively, (d) parametric plot of the Janus flow in the $x,y$ variables, the $\cN=(4,4)$ vacuum is at the origin and the dots denote the locations of the $\cN=(1,1)$ vacua. Blue: Janus between $\cN=(4,4)$ vacua, red: Janus between $\cN=(1,1)$ vacua, green: RG-Janus between $\cN=(4,4)$ and $\cN=(1,1)$. We have set $\alpha=1.2$ for these examples.
} \label{fig-trun2}
\end{figure}

In figure \ref{fig-trun2} we display examples of solutions to the flow equations representing Janus flows between $\cN=(4,4)$ vacua, $\cN=(1,1)$ vacua and RG-flow Janus solutions between  $\cN=(4,4)$ and $\cN=(1,1)$ vacua. We note that the flows involving the $\cN=(1,1)$ vacua are a new feature of the truncation. As discussed in section \ref{opspec} the $\cN=(1,1)$ is a repulsive fixed point of the flow and to obtain the numerical solutions one has to fine tune the initial conditions at the turning point to approach  the $\cN=(1,1)$ vacuum. This implies that choosing an initial $p(0)$, the initial $q(0)$ for which an RG-flow solution exists, is fixed. A third kind of flow solution corresponds to a Janus solution interpolating between $\cN=(1,1)$ vacua, since both vacua are repulsive such solutions only exist for a discrete set of initial conditions. Note that the asymptotic value of $p$ when the $\cN=(4,4)$ vacuum is obtained can take any value and determines the angle with which the point $(x,y)=(0,0)$ is approached  in the parametric $x,y$ plot.
 
\subsection{Truncation 3}

The matrix $A_1$ for this truncation is given by
\begin{align}
A_1= 
\left(
\begin{array}{cccccccc}
  0&0   &0  &{b+c\over 2}&0&{a+d\over 2}&0&0  \\
  0&-{c}    &{-b+c\over 2} &0&{-a+d\over 2}&0&0&d  \\
0 &{-b+c\over 4}   &-{c}  &0&-d&0&0&{a-d\over 2}\\
 {b+c\over 2} & 0  &0  &0&0&0&{-a-d\over 2}&0  \\
 0 &{-a+d\over 2}   &-d  &0&{c} &0&0&{-b+c\over 2}   \\
{a+d\over 2} & 0  &0  &0&0&0&{b+c\over 2}&0  \\
0  &0   &0  &{-a-d\over 2}&0&{b+c\over 2}&0&0  \\
0  &d  & {a-d\over 2} &0&{-b+c\over 2} &0&0&{c}  \\
\end{array}
\right) 
\end{align}
where 
\begin{align}
a&= \sin^3 p  (1+ \alpha \cosh^3 q)\nonumber \\
b&= \cos^3 p  ( \alpha+  \cosh^3 q)\nonumber \\
c&= \sin^2 p \cos p \cosh q  (1+ \alpha \cosh q)\nonumber \\
d&= \cos^2 p \sin p \cosh q  ( \alpha+  \cosh q)
\end{align}

\begin{figure}
\centering
\begin{subfigure}[b]{0.3\textwidth}
	\includegraphics[width=\textwidth]{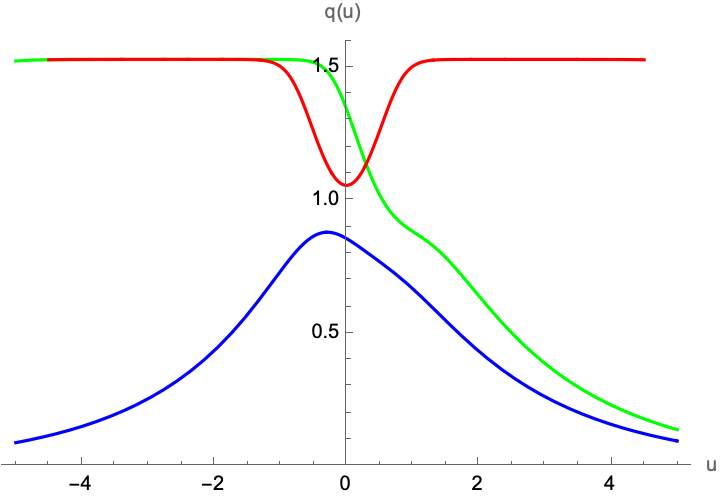}
	\vspace{-0.75cm}
	\caption{}
	\label{fig6-a}
\end{subfigure}
\hspace{1in}
\begin{subfigure}[b]{0.3\textwidth}
	\includegraphics[width=\textwidth]{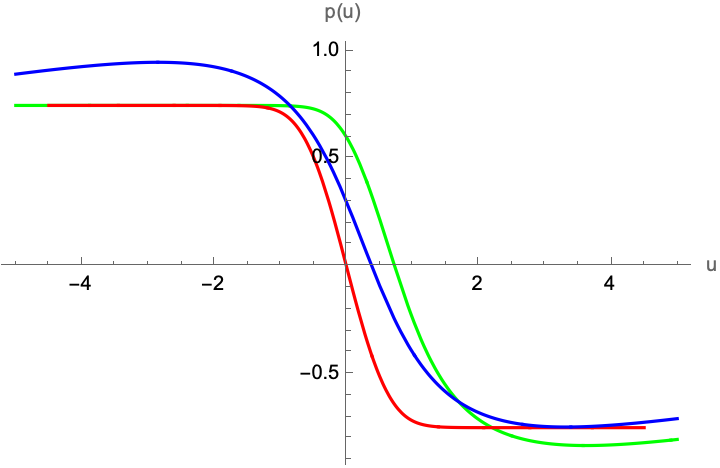}
	\vspace{-0.75cm}
	\caption{}
	\label{fig6-b}
\end{subfigure}
\\
\vspace{0.5cm}
\begin{subfigure}[b]{0.3\textwidth}
	\includegraphics[width=\textwidth]{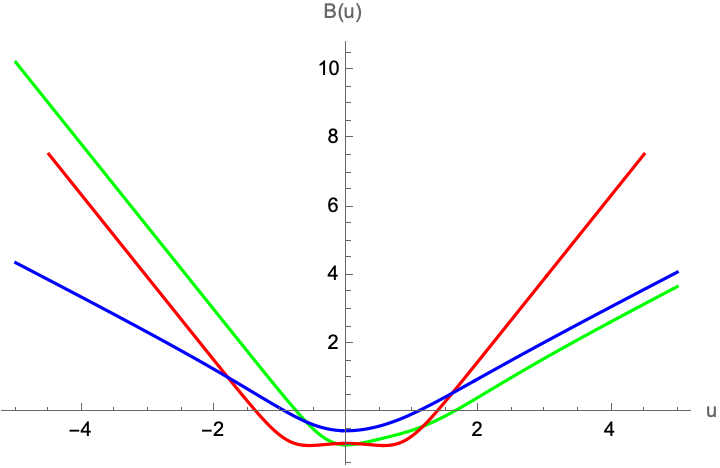}
	\vspace{-0.75cm}
	\caption{}
	\label{fig6-c}
\end{subfigure}
\hspace{1in}
\begin{subfigure}[b]{0.3\textwidth}
	\includegraphics[width=\textwidth]{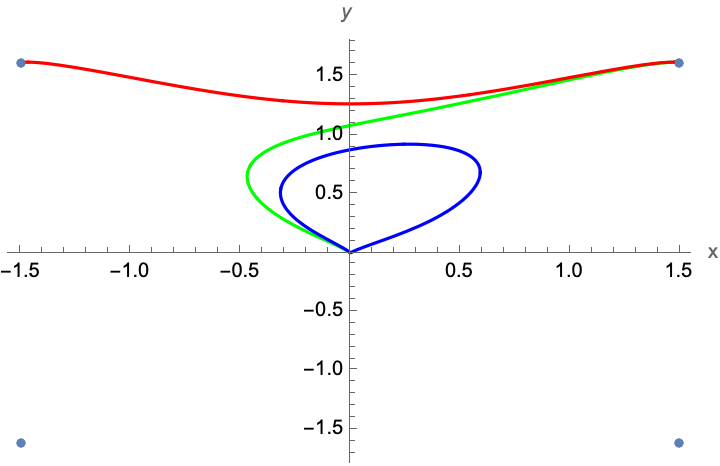}
	\vspace{-0.75cm}
	\caption{}
	\label{fig6-d}
\end{subfigure}
\caption{Truncation 3: (a)-(c) plots of $p,q,B$ respectively, (d) parametric plot of the Janus flow in the $x,y$ variables, the $\cN=(4,4)$ vacuum is at the origin and the dots denote the locations of the $\cN=(1,1)$ vacua.  Blue: Janus between $\cN=(4,4)$ vacua, red: Janus between $\cN=(1,1)$ vacua, green: RG-Janus between $\cN=(4,4)$ and $\cN=(1,1)$. We have set $\alpha=1.2$ for these examples.
} \label{fig-trun3}
\end{figure}

The eigenvectors $n^{(1)}_\pm$ of $A_1$  with eigenvalues $\pm w_0$ corresponding to the unbroken  $\cN=(1,1)$ supersymmetries  are given by
\begin{align}
n^{(1)}_\pm  =
\left(
\begin{array}{c}
0   \\
- {b-3c \pm 2 w_0 \over a-3d} \\
  {b-3c \pm 2 w_0 \over a-3d}\\
0 \\
 1\\
 0\\
 0\\
 1
\end{array}
\right), \quad w_0= {1\over 2} \sqrt{ (a-3d)^2 + (b-3c)^2}
\end{align}
The rest of the eigenvectors of $A_1$ do not have eigenvalues which satisfy the supersymmetry condition (\ref{susyvac}) for the $\cN=(1,1)$ vacuum. Note that all of them reduce to the ones of truncation 1 for the $\cN=(4,4)$ vacuum.

In figure \ref{fig-trun3} we display a sample of solutions to the flow equations representing Janus flows between $\cN=(4,4)$ vacua, $\cN=(1,1)$ vacua and RG-flow Janus solutions between an $\cN=(4,4)$ and $\cN=(1,1)$ vacuum. We note that the solutions behave qualitatively similar to the ones displayed for truncation 2.

%%%%%%%%%%%%%%%%
\section{Holographic calculations}\label{sec5}
%%%%%%%%%%%%%%%%

In this section we will perform some holographic calculations for the solutions obtained in the section \ref{sec4}.  In particular we will calculate the masses for the fluctuations of the scalar fields around the $\cN=(4,4)$ and $\cN=(1,1)$ vacua. This will allow us to identify the dimensions of the dual operators which are turned on in the flows. One of the results is that for truncation 2 and 3 the mass squared of the fluctuations are positive, corresponding to operators with scaling dimensions $\Delta>2$. Since the behavior near the AdS vacuum is given by 
\begin{align}\label{phiexp}
\lim_{\epsilon\to 0}  \phi \sim \bar \phi + c_1 \epsilon^\Delta +c_2 \epsilon^{2-\Delta} +\cdots
\end{align}
where $\epsilon\to 0$ corresponds to approaching the AdS boundary,  the initial conditions have to be fine tuned in order to make the repulsive term $c_2   \epsilon^{2-\Delta}$ very small.  In addition we consider the entanglement entropy of a symmetric region around the defect  \cite{Azeyanagi:2007qj,Chiodaroli:2010ur,Jensen:2013lxa,Estes:2014hka,Gutperle:2015hcv} and give a prescription to obtain the defect entropy (or g-factor) \cite{Affleck:1991tk}.

\subsection{Operator spectrum}
\label{opspec}

The $\cN=(4,4)$ vacuum has  $q_i=0, i =1,2,3,4$. Since the kinetic terms for $p_i$ are vanishing the $x_i,y_i$ defined in (\ref{xyqp}) are better suited to analyze the fluctuations. Expanding around the $x_i=y_i=0$ vacuum one finds  for the quadratic term of the fluctuations,
\begin{align}
{1\over e} {\cal L}_{(2)} =  {1\over 4} \sum_i  \Big( \partial_ \mu \delta x_i \partial^\mu \delta x_i +  \partial_ \mu \delta y_i \partial^\mu \delta y_i \Big)+{g_c^2 \over 4} \sum_i \Big( (1+2\alpha )\delta x_i^2 + \alpha(\alpha+2)\delta y_i^2\Big)
\end{align}
from which we can read off the masses of the scalar fluctuations.  Then the masses determine the conformal scaling dimensions
\begin{align}\label{scaldim}
\Delta_{\pm}= 1\pm \sqrt{1+ m^2 R^2}
\end{align}
where $R$ is the AdS radius of the vacuum.
 Setting $g_c=1/(1+\alpha)$ to obtain a unit radius $AdS_3$ for the $\cN=(4,4)$ vacuum and the  standard AdS/CFT relation the conformal dimensions of the dual operators are displayed in table \ref{tab:one}.
\begin{table}[htp]
\begin{center}
\begin{tabular}{|c|c|c|c|}
\hline
$\cN=(4,4)$&$m^2$&$ \Delta_+$& $\Delta_-$\\ 
\hline
$\delta x_i$ &$ -{1+ 2\alpha \over (1+\alpha)^2}$&${1+2\alpha\over 1+\alpha}$&${1\over 1+\alpha}$ \\
\hline
$\delta y_i$ &$ -{\alpha( 2+\alpha)\over (1+\alpha)^2}$&${2+\alpha\over 1+\alpha}$&${\alpha\over 1+\alpha} $\\
\hline
\end{tabular}
\end{center}
\caption{Mass and conformal dimensions of scalar fluctuations for the $\cN=(4,4)$ vacuum}
\label{tab:one}
\end{table}%
Note that $\Delta_+$ gives the scaling dimension of the dual operator in  the standard quantization which  takes values between $1< \Delta_+<2$ for $\alpha>0$, whereas $\Delta_-$ corresponds to the alternative  quantization and $0< \Delta_-<1$ for $\alpha>0$. Supersymmetric flows are related to the standard quantization which we will adapt in the following \cite{Bianchi:2001de}. We note that the $\cN=(4,4)$ vacuum is attractive since  both $x$ and $y$ are dual to operators with $ \Delta<2$ and the initial conditions do not have to be fine tuned for (\ref{phiexp}) to approach the vacuum value.

\begin{figure}
\centering
\begin{subfigure}[b]{0.37\textwidth}
	\includegraphics[width=\textwidth]{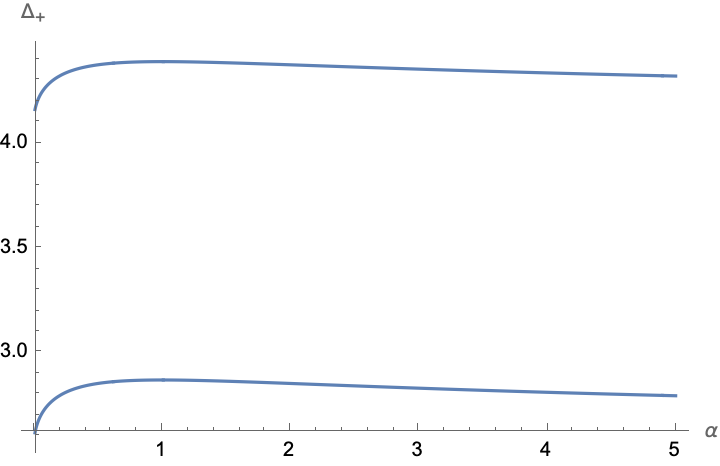}
	\vspace{-0.75cm}
	\caption{}
	\label{fig-delplot2}
\end{subfigure}
\hspace{.2in}
\begin{subfigure}[b]{0.37\textwidth}
	\includegraphics[width=\textwidth]{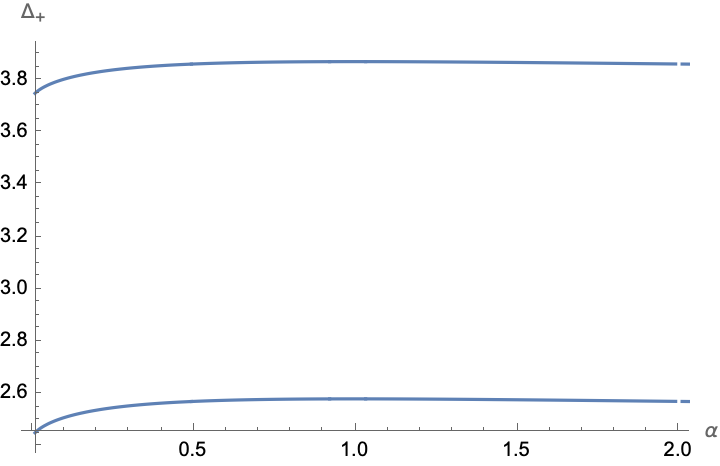}
	\vspace{-0.75cm}
	\caption{}
	\label{fig-delplot3}
\end{subfigure}
\caption{(a)  Conformal dimension of operator dual to scalar fluctuations around the $\cN=(1,1)$ vacuum for truncation 2, (b) same for truncation 3.} \label{fig-plotdel}
\end{figure}

For truncations 2 and 3 we can determine the scaling dimensions of the operators at the $\cN=(1,1)$ vacuum by expanding  the scalar action around the vacuum  to second order and diagonalizing the resulting scalar Lagrangian. The resulting expressions are quite unwieldy and we present the plots of the scaling dimensions of the two modes as a function of $\alpha$ in figure \ref{fig-plotdel}. We note that the scaling dimensions are larger than 2 and hence the $\cN=(1,1)$ corresponds to a repulsive fixed point and the initial conditions have to be fine-tuned.

\subsection{Holographic entanglement entropy}

\begin{figure}
\centering
 \begin{subfigure}[b]{0.42\textwidth}
	\includegraphics[width=\textwidth]{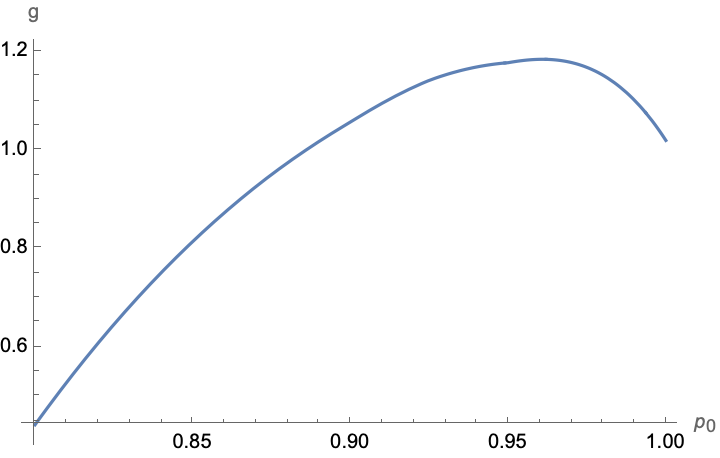}
	\vspace{-0.75cm}
	\caption{}
	\label{fig-delplot-2}
\end{subfigure}
\hspace{.2in}
\begin{subfigure}[b]{0.45\textwidth}
	\includegraphics[width=\textwidth]{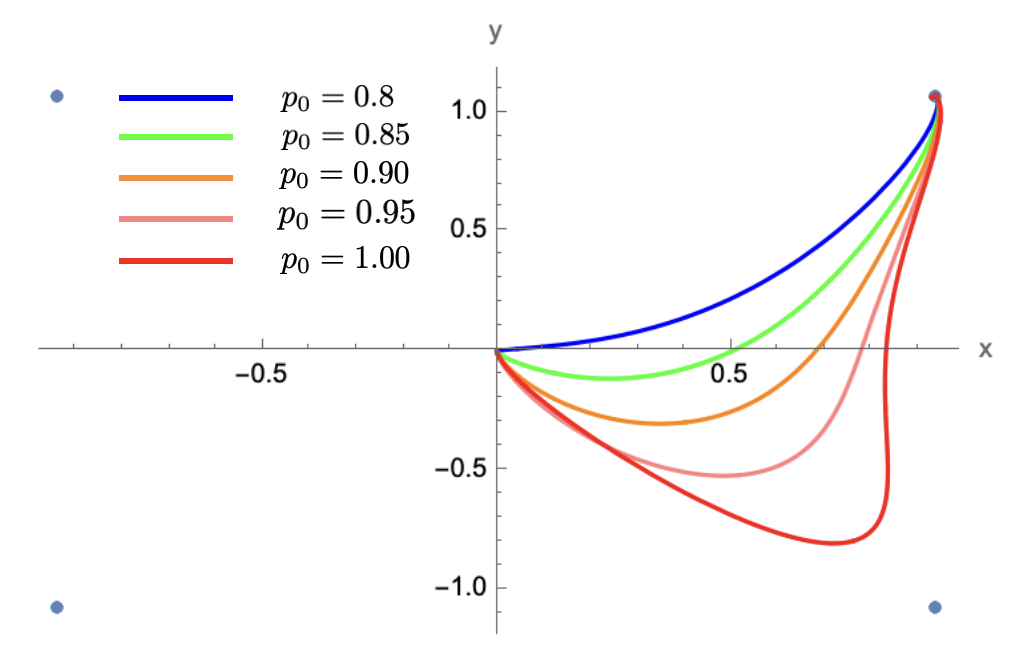}
	\vspace{-0.75cm}
	\caption{}
	\label{fig-delplot-3b}
\end{subfigure}
\caption{(a) Plot of boundary entropy for RG-flow interface for truncation 2, as a function of initial condition $p_0$ at the turning point for $\alpha=1.4$. (b) Illustration of RG-flows for some initial values of $p_0$.} \label{fig-plotg}
 \end{figure}

The Ryu-Takayanagi prescription  \cite{Ryu:2006bv} relates holographic entanglement entropy to the area of a minimal surface in the bulk which when approaching the $AdS$ boundary ends  at the border of the entangling surface. For a three dimensional static bulk spacetime this corresponds to a geodesic in the bulk which terminates at the ends of the entangling interval on the boundary.  For the $AdS_2$ sliced metric (\ref{janusan}) and an entangling surface which is symmetric about the defect and of length $2L$,  such a geodesic is simply parameterized by $u$ and constant $z=L$. The entanglement entropy is then given by
\begin{align}\label{seeform}
S_{EE}(L) = {1\over 4 G_N} \int_{u_{-\infty}}^{u_{+\infty}} du=  {1\over 4 G_N} \big( u_{-\infty} - u_{+\infty}\big)
\end{align}
where $u_{\pm \infty}$ will be related to an UV Fefferman-Graham cutoff in the following. We will generalize the derivation of \cite{Azeyanagi:2007qj,Chiodaroli:2010ur} to the case of an RG-flow interface where the AdS radius and hence the central charge take different values on both sides of the interface. The asymptotic behavior of the metric is determined by the metric function $B(u)$ as $u\to \pm \infty$
\begin{align}\label{bfunlim}
\lim_{u\to \pm \infty} B(u) = \pm {u\over R_{\pm}} + \ln \lambda_{\pm} - \ln 2+ o({1\over u})
\end{align}
In the two asymptotic regions we can define a Fefferman-Graham coordinate system by defining a new coordinates $\hat u_{\pm}$
\begin{align}
u\to \pm \infty:  \quad u &= R_{\pm}  \hat u_{\pm} \mp R_{\pm} \ln \lambda_{\pm}+ o({1\over u})
\end{align}
and then the coordinates $\zeta_{\pm}, \eta$
\begin{align}
u\to +\infty:&\quad\quad   e^{-2\hat u_+}  ={1\over 4}  {\zeta_+^2\over\eta^2} + o(\zeta_+^4), \quad \quad z = \eta \Big( 1+{1\over 2} {\zeta_+^2\over \eta^2}\Big)   + o(\zeta_+^4) \nonumber \\
u\to -\infty:&\quad\quad \;  e^{2\hat u_-}  ={1\over 4}  {\zeta_-^2\over\eta^2} + o(\zeta_-^4), \quad \quad z =- \eta \Big( 1+{1\over 2} {\zeta_-^2\over \eta^2}\Big)   + o(\zeta_-^4) \nonumber \\
\end{align}
This expansion is valid for $\eta >> \zeta_{\pm}$, i.e. if we consider an entanglement interval which is far away from the interface. In this limit the metric becomes
\begin{align} \label{fgrelate}
\zeta_{\pm} \to 0& \quad \quad ds^2 = R_{\pm}^2 \Big( { - d\zeta_{\pm}^2 -d\eta^2 + dt^2 \over \zeta_{\pm }^2}\Big)+ o(1)
\end{align}
It follows that $R_\pm$ defined in (\ref{bfunlim}), corresponds to the asymptotic AdS radius  and the left and right side of the interface respectively and a Fefferman-Graham cutoff is given  by setting $\zeta_{\pm}=\epsilon$. For the entanglement region located at $z=L$ in follows from (\ref{fgrelate}) that the FG cutoff is related to the $u_\pm$ cutoff as follows
\begin{align}
u_\pm =\mp R_{\pm \infty } \ln \left({1\over 2} {\epsilon \over L} \right)\mp R_\pm\ln \lambda _\pm
\end{align}
Plugging this into (\ref{seeform}) gives the entanglement entropy
\begin{align}\label{seeres}
S_{EE}&= {1\over 4 G_N}  \Big( (R_+ +R_-) \ln {L\over  2 \epsilon}  - R_+ \ln \lambda_+ - R_- \ln \lambda_-\Big) \nonumber \\
&= {c_++c_-\over 6}   \ln {L\over  2 \epsilon } -  {c_+ \over 6}   \ln \lambda_+-  {c_- \over 6}   \ln \lambda_-+ o(\epsilon)
\end{align}
the constant term gives the boundary entropy
\begin{align}
g= -  {c_+ \over 6}   \ln \lambda_+-  {c_- \over 6}   \ln \lambda_-
\end{align}
Where $c_\pm$ is the central charge for the two CFTs on either side of the RG interface. 
The g-factor is given by the second and third term in (\ref{seeres}).
For a Janus  interface we have $c_+=c_-=c$, whereas the central charges differ on both sides of the interface for a RG-flow interface.
It is straightforward to determine the $R_\pm$ and $\ln \lambda_\pm$ by numerically fitting the metric functions (see plot (c) in figures \ref{fig-trun2} and \ref{fig-trun3}) to determine the slope and the intercept (\ref{bfunlim}) in the limit of large $|u|$. We will give an example of numerical results by presenting the g-factors as for the RG-interface between the $\cN=(4,4)$ vacuum and a $\cN=(1,1)$ vacuum in truncation 2. As discussed in section \ref{sec4:trun2} there exists a unique RG-flow interface for a choice on initial condition $p_0$. In figure \ref{fig-plotg} we present the g-factor as a function of the initial condition for a particular value of $\alpha= 1.4$.

%%%%%%%%%%%%%%%%
\section{Discussion}\label{sec6}
%%%%%%%%%%%%%%%%
 				
In this paper we found holographic interface solutions in three dimensional gauged supergravity theories. An important feature of these theories is that they have AdS vacua which preserve $\cN=(1,1)$ supersymmetry in addition to the $\cN=(4,4)$ AdS vacuum.  This feature allows us to find solutions which correspond to interfaces between two $\cN=(4,4)$ vacua on both side, $\cN=(1,1)$ on both sides, as well as RG-flow interfaces which have a $\cN=(4,4)$ on one side and $\cN=(1,1)$ vacuum on the other. 
We derived BPS flow equations which are three first order nonlinear differential equations for the two scalars $p,q$ which are  non zero  in the truncations as well as the warp factor $B$ of the $AdS_2$ slicing. By using the freedom to shift the warping coordinate $u$ by a constant we can choose the initial conditions for the flow as the value of $p$ and $q$ at the turning point of the warp factor, where $B'=0$. 
 In fact we use the BPS equations to determine the initial conditions for the second order equation motion following from the variation of the action. The numerical accuracy of the solution is tested by checking the BPS equations away from the point where the initial conditions were fixed. 	

The $\cN=(1,1)$ extrema are repulsive fixed points of the flow and hence the initial condition have to be fine tuned using a shooting method. This is possible by fixing one scalar initial condition and varying the other in order to come closer and closer to the $\cN=(1,1)$ vacuum in the flow. Our results indicate that the qualitative behavior of the solutions for general $\alpha$ is quite similar to the behavior of the $\alpha=1$ solutions obtained in \cite{Chen:2021mtn}.  In addition we have considered the entanglement entropy for the Janus and RG-flow solutions. Since for the RG-flow solutions the central charges and hence AdS radii are different on both sides of the interface one has to carefully consider the UV cut-off. It is possible to determine the g-function or interface entropy from the numerical solution by a linear fit of the warp factor $B$. 

We have considered truncations of the scalars to two nonzero scalars $q$ and $p$ (or $x$ and $y$), it would be interesting to generalize this since it would then be possible to consider more complicated flows between different $\cN=(1,1)$ vacua. It would also be interesting to investigate the solutions we have found can be lifted and have a representation in $\AdS_3\times S^3\times S^3 \times S^1$ holography. It would also be interesting to see whether the prescription for the interface entropy can be applied to other examples of RG-flow interfaces.  We leave these interesting questions for future work.

\section*{Acknowledgements}

The work of M.~G.~was supported, in part, by the National Science Foundation under grant PHY-19-14412.  The authors are grateful to Kevin Chen for initial  collaboration.
The authors are grateful to the Mani L.~Bhaumik Institute for Theoretical Physics for support.

\newpage
\appendix

%%%%%%%%%%%%%%%%
\section{$\SO(8)$ Gamma matrices}\label{appendix-gamma}
%%%%%%%%%%%%%%%%

The formulation of the $N=8$ three dimensional gauged supergravity utilizes  with $8 \times 8$ Gamma matrices $\Gamma^I_{A \dot A}$ and their transposes $\Gamma^I_{\dot A A}$, they satisfy the Clifford algebra,
	\begin{align} \Gamma^I_{A \dot A} \Gamma^J_{\dot A B} + \Gamma^J_{A \dot A} \Gamma^I_{\dot A B} = 2 \delta^{IJ} \delta_{AB}
	\end{align}
Explicitly, we use the basis as given in Green-Schwarz-Witten \cite{Green:1987sp},
	\begin{align}
	\Gamma^8_{A \dot A} &= 1 \otimes 1 \otimes 1~, &  \Gamma^1_{A \dot A} &= i \sigma_2 \otimes i \sigma_2 \otimes i \sigma_2  \nonumber \\
	\Gamma^2_{A \dot A} &= 1 \otimes \sigma_1 \otimes i\sigma_2~, &  \Gamma^3_{A \dot A} &= 1 \otimes \sigma_3 \otimes i\sigma_2 \nonumber \\
	\Gamma^4_{A \dot A} &= \sigma_1 \otimes i\sigma_2 \otimes 1~, &  \Gamma^5_{A \dot A} &= \sigma_3 \otimes i\sigma_2 \otimes 1  \nonumber \\
	\Gamma^6_{A \dot A} &= i\sigma_2 \otimes 1 \otimes \sigma_1~, &  \Gamma^7_{A \dot A} &= i\sigma_2 \otimes 1 \otimes \sigma_3
	\end{align}
The matrices $\Gamma^{IJ}_{AB}$, $\Gamma^{IJ}_{\dot A \dot B}$ and similar are defined as antisymmetrized products of $\Gamma$s with the appropriate indices contracted.
For example,
	\begin{align} \Gamma^{IJ}_{AB} \equiv \frac{1}{2} (\Gamma^I_{A \dot A} \Gamma^J_{\dot A B} - \Gamma^J_{A \dot A} \Gamma^I_{\dot A B}) 
	\end{align}

\newpage
	
\providecommand{\href}[2]{#2}\begingroup\raggedright\endgroup

\end{document}